# Multidimensional photonic computing

Ivonne Bente[1,*], Shabnam Taheriniya[2,*], Francesco Lenzini[3], Frank Brückerhoff-Plückelmann[1,2], Michael Kues[4], Harish Bhaskaran[5], C David Wright[6] and Wolfram Pernice[1,2]

[1]*Physics Institute, University of Münster, 48149 Münster, Germany*
[2]*Kirchhoff-Institute for Physics, Heidelberg University, 69120 Heidelberg, Germany*
[3]*Consiglio Nazionale delle Ricerche, Istituto di Fotonica e Nanotecnologie (CNR-IFN), 20133 Milan, Italy*
[4]*Institute for Photonics, Leibniz University Hannover, 30167 Hannover, Germany*
[5]*Department of Materials, University of Oxford, Parks Road, Oxford, OX1 3PH, UK*
[6]*Department of Engineering, University of Exeter, Exeter, EX4 4QF, UK*

*These authors contributed equally to this work.

**The rapidly increasing demands for computational throughput, bandwidth, and memory capacity fueled by breakthroughs in machine learning pose substantial challenges for conventional electronic computing platforms. For digital scaling to keep pace with the accelerating growth of artificial intelligence (AI) models beyond the trajectory of Moore's law, computational power has to double roughly every three months. Historically, advancing compute performance relied on spatial scaling to increase the transistor count on a given chip area and, more recently, the development of parallel and multi-core architectures. Exponential scaling on trajectories much steeper than what can be achieved by such conventional strategies, and in line with the demands of AI, can be achieved with computing platforms that process data using multiple, orthogonal dimensions available to photons. Here we elucidate pivotal developments in the realization of multidimensional computing platforms based on photonic systems. Moving to such architectures holds enormous promise for low-latency, high-bandwidth information processing at reduced energy consumption.**

# 1. Introduction

The growing use of AI and Machine Learning (ML) technology is pushing the limits of core computing with respect to efficient real-time data processing and low latency communication. As the computing landscape evolves to cater to the needs of an increasingly AI-enhanced society, the viability of future tech relies on incorporating sustainable and environmentally conscious alternatives. At present, the demand for higher computing power to train and apply advanced AI models in Computer Vision, Natural Language Processing, and Speech Learning has by far exceeded Moore's Law scaling[1,2]. Large language models with billions of model parameters, such as OpenAI's GPT, Meta's Llama and Google's Gemini systems, present severe challenges on available computational resources[3–5]. This has widened the gap between the available computational capacity and the rapidly increasing volume of collected data needing processing. At the same time, meeting computational challenges leads to steadily increasing power consumption and carbon emissions[6,7], which are at odds with the goals to achieve carbon neutrality in the coming decades.

Given the current technological landscape, two prominent directions emerge to address future scaling needs in terms of hardware: firstly, the development of specialized accelerators specifically designed to efficiently process large datasets characterized by their sheer volume rather than algorithmic complexity[8–11]. And secondly, specialized computing hardware directed at complex and computationally intensive tasks, such as simulating molecular and atomic interaction[12–15], cryptographic algorithms[16–21], climate[22,23] or financial modeling and risk analysis[24–28], yet with moderate size in the underlying data sets. In the first case, accelerators designed to efficiently execute arithmetic operations such as matrix-vector multiplications in graphics processing units (GPUs), tensor processing units (TPUs) and neural processing units (NPUs) complement traditional central processing units (CPUs) in AI workloads. In the latter case where non-AI complex algorithms are executed, non-classical machines exploiting quantum principles for physical computing are handled as promising architectures due to their inherent capacity to perform complex computations which are intractable on classical hardware. Currently, access to computing hardware with long-term accelerated scaling is limited in our tech environment. In order to satisfy the computational needs of future AI, ML and advanced quantum algorithms, a concentrated effort is needed to move beyond current scaling options.

Shifting focus from today's predominantly employed Fermionic information carriers in electrical circuits to Bosonic information carriers, in particular photons, allows for accelerated computational scaling. This is achieved by harnessing photons' ability for superposition within a state and exploiting orthogonal degrees of freedom for data representation. Through this, a single 'Bosonic chip' can significantly enhance computational density, by enabling multidimensional computing. This concept is exploited in high-speed optical communication through techniques like wavelength division multiplexing (WDM)[29–31], in-phase and quadrature (IQ) modulation[32–34], and polarization[35] as well as spatial mode encoding[29], all of which contribute to highly parallelized signal processing. Exploiting different degrees of freedom in computing can similarly lead to the development of highly parallelized architectures in both classical and non-classical (quantum) applications, where computing involves creating high-dimensional entanglement between multi-level systems. One of the advantages of photonic platforms over other quantum systems, such as ions or superconducting devices, lies in the enhanced quantum information capacity and noise resilience of high-dimensional systems, which can, for example, be generated by exploiting spectral and time degrees of freedom in pulsed frequency combs[36]. Improved scalability and the potential for implementing error correction strategies resides on the combination of multiple degrees of freedom to form hyper-entangled states, supported by the unique characteristics of bosonic systems. Along the same lines, in classical brain-inspired or neuromorphic photonic computing strategies, incorporating spectral, polarization, and mode degrees of freedom at higher intensities enables dramatic increase of the computational density of a given chip area, despite the large footprint of photonic devices compared to state-of-the art digital transistors.

Here, we offer our viewpoint on the advancement of multidimensional computing, with emphasis on utilizing photonic principles for brain-inspired and quantum technologies. Our perspective centers on utilizing orthogonal degrees of freedom within a single computational structure, with focus on achieving exponential scaling in computational power.

**Box 1: Multidimensional photonic computing**

As computing demands continue to grow, there will be a need for new hardware that facilitates the handling of significantly complex tasks through unconventional ways of information processing[37]. To this end, architectures that deviate from the traditional von Neumann approach are becoming more prominent. This includes distributed multilayer processing models inspired by a concept that is neuromorphic in form[38] and provides access to new levels of speed and efficiency[39]. In addition, a range of optimized solutions for AI applications has been developed, including hyperdimensional computing[40–42] and hybrid models that integrate optical and electronic analog computing technologies[11]. In parallel, to accommodate the increasing complexity in algorithms intractable on classical hardware, quantum computers are envisaged to be integrated with digital platforms that enable long-term data storage and conventional information processing[43,44].

A common approach in standard electronic systems for enhancing computational speed and managing larger data sets is to use highly parallel data processing in specialized accelerators and multi-core architectures. Quantum processors, however, excel at solving complex problems beyond the reach of classical computers by operating in a physical Hilbert space, which expands exponentially with the addition of physical qubits.

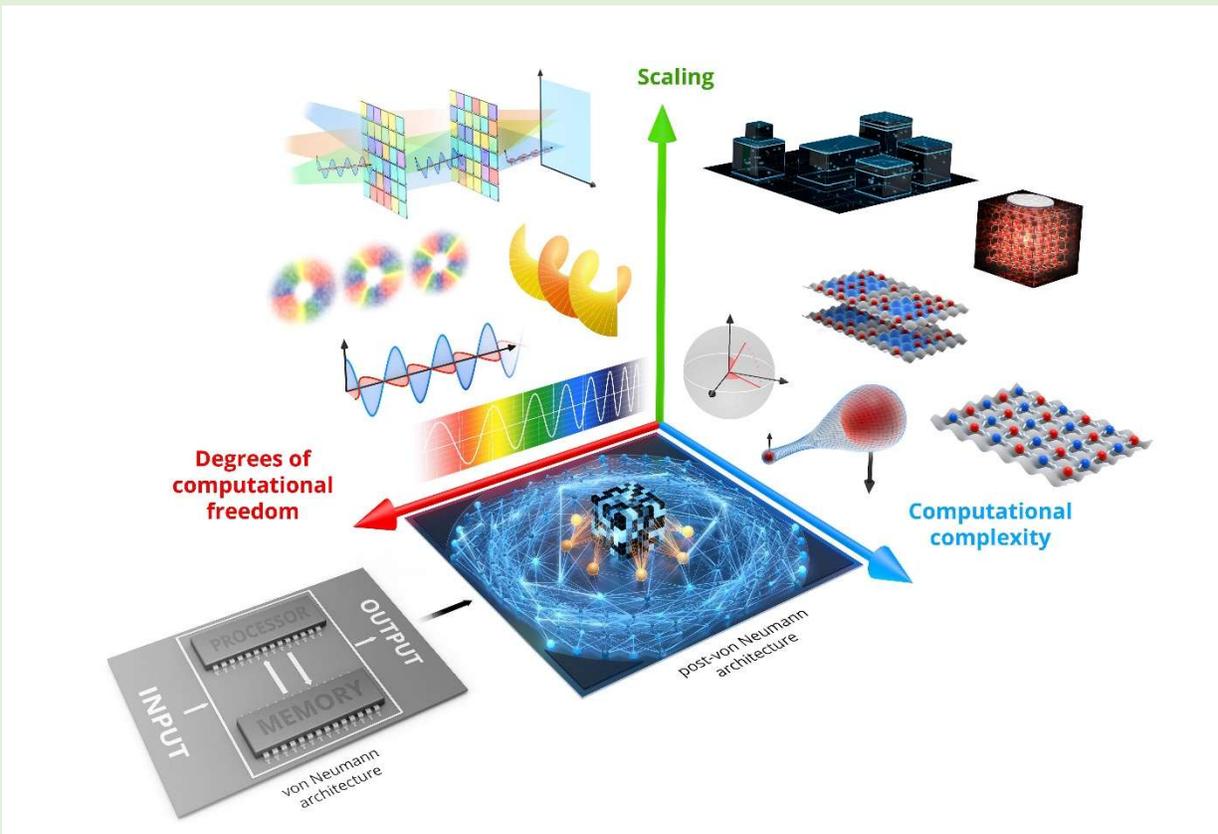

**Figure B1: Multidimensional photonic computation by using multi-parameter scaling to move beyond current von-Neumann architectures.** Orthogonal degrees of freedom available to Bosonic information carriers enable increased computational density for a given chip area.

> In the optical domain, using photons as information carriers with orthogonal degrees of freedom for data encoding in optical neural networks and optical quantum computers represent a shift towards post-von-Neumann computing. These systems exploit the unique advantages of the optical domain, particularly the ability to combine various forms of multiplexing, including wavelength, polarization, spatial modes and temporal degrees of freedom. Multiplexing enhances computational performance without increasing the chip area, unlike traditional electronic (fermionic) computing which is reaching the physical limits of Moore's Law [45]. Transitioning from binary/digital to physical analog computing, including photonic computing, results in new challenges to reduce noise and to employ statistical methods.

## 2. Data heavy versus complexity-driven architectures

Next-generation photonic platforms hold promise for changing the landscape of classical and quantum computing for data-heavy and complexity-heavy applications, utilizing light's unique properties to process information with unparalleled speed and efficiency. While having achieved notable data throughput levels, classical computing built on electronic circuits is constrained by fundamental physical and efficiency limitations. With the push towards future exascale computing architectures, ultrahigh throughput is expected to become a critical component.

In non-quantum data heavy information processing, using photons for implementing arithmetic operations in analog computing systems enables single-shot operation at very high bandwidth and throughput. Mid- and long-term scalability can be made possible through the application of linear and non-linear optical techniques, which allow brain-inspired or neuromorphic systems to emulate the analog computation functions of the nervous tissue. Especially intriguing is the combination of in-memory computing-based neuromorphic architectures with orthogonal degrees of freedom for data encoding.

In contrast, for complexity-driven architectures hyperdimensional computing and quantum computing approaches offer viable scaling routes. Quantum computing capitalizes on the principles of quantum mechanics to execute intricate computations more effectively than classical systems, particularly in cryptography and complex simulation tasks. However, state-of-the-art quantum processors currently struggle with managing high data throughput due to error rates, limited coherence times, and the fragile nature of qubit operations, which pose significant challenges to data and signal reliability in large-scale applications. The seamless integration of these two computing paradigms on a photonic platform offers a promising path forward, aimed at maximizing data processing capabilities while minimizing environmental interaction and reducing thermal noise.

## 2.1. Data heavy computation

Enhancing computing capabilities through the use of photonic concepts has led to substantial progress in neuromorphic architectures. Sludds et al. discuss an optical computing network that leverages Wavelength Division Multiplexing (WDM) over a 3 THz optical bandwidth to efficiently perform image recognition across long distances (86 km) with minimal power consumption[46]. This design incorporates edge computing elements, employing smart transceivers and time-domain multiplexing. Xu et al. introduce a photonic convolutional accelerator that uses a combination of time, wavelength, and spatial multiplexing[8] to reach a processing speed of 11 tera-OPS. Their method encodes convolutional kernel weights into microcomb lines as optical power, simplifying data processing by transmitting temporal input data through kernel wavelength channels and combining the optical signals at a photodetector (cf. Fig. 1a).

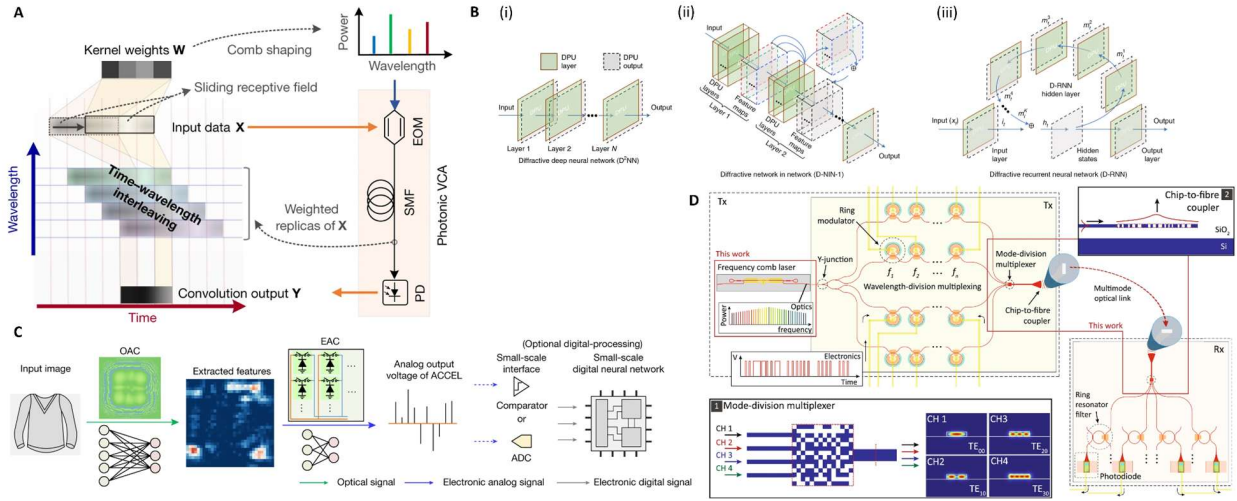

**Figure 1: Overview of neuromorphic photonic computing approaches and platforms.** A) Schematic image of the combination of the photonic convolutional accelerator, which makes use of temporal and wavelength dimensions to obtain high computational throughput of up to 11 TOPS[8]. In the scope of this work, another arm consisting of an EOM, SMF, EDFA and a demultiplexer are added to also make use of the spatial dimensions. B) The diffractive optoelectronic processor is shown in different configurations where it works as deep neural network, network in network, or recurrent neural network[50]. C) Scheme of the processing flow of a photo-electronic chip combined with an electronic analog signal processor[11]. The input image is processed by the OAC extracting small-scale features which are then further processed by the EAC giving the final result. This can be further processed by an optional digital unit. D) An inverse-designed mode multiplexer enables the combination of wavelength and mode division multiplexing[31].

In related work, Tan et al. developed a video image processor using a Kerr microcomb and multiplexing techniques to achieve a high processing bandwidth of 17.4 Tbit/s[9]. These efforts share the goal of projecting peta-OPS speeds and expanding the functional range by incorporating polarization and additional spatial dimensions.

Focusing on neural network advancements and optimization solutions, several studies aim to enhance the computing capabilities of photonic technologies. Feldmann et al. applied WDM technology to design scalable circuits for photonic neural networks, simulating the behavior of basic integrate-and-fire neurons[47]. In another approach, Feldmann et al. incorporated on-chip frequency combs and photonic tensor cores, achieving 2 Tera-MAC operations per second in a specialized application-specific processor[48]. Researchers from Microsoft and Cambridge introduced an Analog Iterative Machine (AIM) that employs spatial-division multiplexing[49]. AIM reportedly surpasses existing quantum hardware in solving optimization problems at room temperature with unprecedented accuracy. Zhou et al. present a reconfigurable diffractive processing unit that uses spatial modulation and temporal multiplexing, supporting millions of neurons[50]. This unit can adapt to various types of neural networks, such as diffractive deep neural networks, diffractive networks in networks, and diffractive recurrent neural networks (cf. Fig. 1b). Utilizing spatial mode multiplexing, Antonik et al. propose a large-scale brain-inspired photonic computer for classifying video-based human actions[51]. Their system modulates the phase of a spatially extended planar wave using a spatial light modulator (SLM). These multiplexing techniques, integrated with advanced photonic elements, aim to surpass classical computers in neural networks for AI and optimization problems.

Chen et al. demonstrate a novel approach in Optical Neural Networks (ONNs) with their compact VCSEL architecture, which utilizes time multiplexing to model systems with tens of billions of neurons[52]. Their unique coherent homodyne multiplication method allows for direct hardware execution of phase-encoded detection-based optical non-linearity, a task typically handled by software in traditional ONNs. In efforts to improve the speed and scalability of photonic neural networks, Zhang et al. present a large-scale photonic Convolutional Neural Network (CNN) employing time-division multiplexing, while also suggesting the use of Spatial Division Multiplexing (SDM) and Wavelength Division Multiplexing (WDM) to significantly enhance system throughput[53]. Their network design is built on Distributed Feedback Laser Diode (DFB-LD) neurons, which process both excitatory and inhibitory inputs. Tait et al. further advance this by integrating WDM on a silicon photonic chip using a microring weight bank approach[54], closely resembling a continuous-time recurrent neural network (RNN) to replicate its behavior in a photonic system.

Chen et al. also propose an all-analog chip combining electronic and optical computing, called ACCEL[11], which significantly reduces energy consumption by optimizing the opto-electronic processing interface. The ACCEL chip integrates Optical Analog Computing (OAC) and Electronic Analog Computing (EAC), with OAC managing data processing and feature extraction, while EAC handles the final result calculations. This architecture (Fig. 1c) supports digital post-processing and achieves an energy reduction by three orders of magnitude.

Further optimization of ONNs can be achieved with algorithms like Physics-Aware Training (PAT)[55], which applies a backpropagation framework to controllable physical systems. PAT is particularly advantageous for inference tasks, especially when there is no clear mathematical isomorphism between the physical layer and conventional artificial neural networks. The key benefit of PAT lies in its ability to simulate the backward pass while the physical system processes the forward pass simultaneously.

Neuromorphic configurations, employing both linear and non-linear optical techniques, imitate the functions of analog computation, pushing the limits of data processing speed. Significant advancements include the integration of microcomb technology with an inverse-designed silicon photonic mode-division multiplexer[31], facilitating error-free transmission of 1.12 Tb/s through both mode- and wavelength-division multiplexing (Fig. 1d). Additionally, a single microcomb ring can achieve an optical data transmission rate of 1.84 Pbit/s by incorporating both spatial and wavelength multiplexing[29]. These photonic neuromorphic architectures, combining high-speed data transmission with formidable computational capabilities, are crucial for the future of exascale computing and underscore the importance of ultra-high throughput.

## 2.2 Complexity-driven computation

Alongside quantum computing approaches, advanced machine learning and computational models that effectively realize abstract mappings and manage complex data structures, employ the concept of hyperdimensional computing (HDC). By focusing on encoding information in hyperdimensional vectors and employing content-addressable memory platforms, HDC significantly improves the support for advanced machine learning algorithms including support vector machines. HDC is particularly notable within computational architectures due to its inherent robustness against uncertainties, often introduced by hardware, making HDC particularly suited to analog systems. For example, by using an electronic in-memory approach, the encoder and associative memory demonstrated by Karunaratne et al. outperform 65 nm CMOS technology in terms of energy efficiency[40]. In another development, Karunaratne et al. introduced a novel attention mechanism that stores unrelated items in key memory as uncorrelated hyperdimensional vectors, thereby enhancing the controller's performance (Fig. 2a) [41]. This advancement enables the application of HDC in CNN controllers. Additionally, Hersche et al. proposed a neuro-vector-symbolic architecture (NVSA), which combines deep neural networks with vector-symbolic architectures (VSAs)[42]. This integration creates a synergistic system that co-designs a visual perception frontend and a probabilistic reasoning backend, leading to improvements in the overall computational architecture (Fig. 2b).

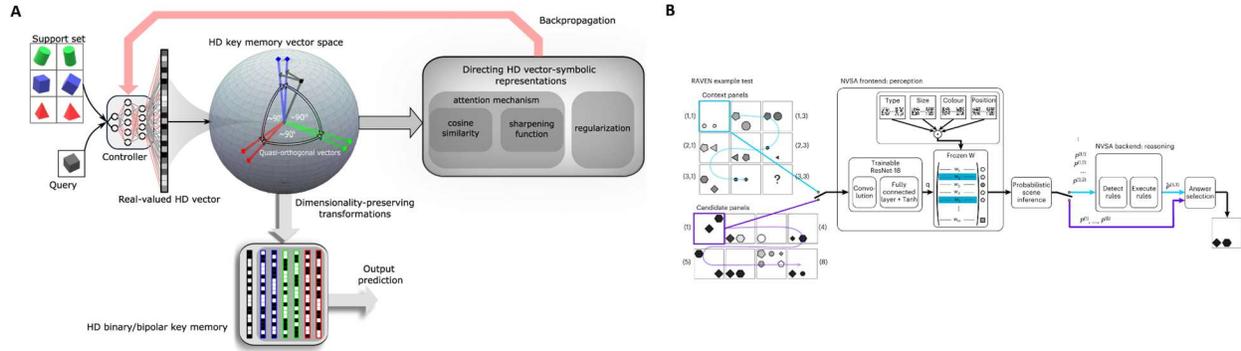

**Figure 2: Highly entangled photonic quantum computation and hyperdimensional computational approaches.** A) Example of high-dimensional (HD) computing where deep network representations are merged with vector-symbolic representations[41]. B) Schematic of a neuro-vector-symbolic architecture (NVSA) for solving Raven's progressive matrices[42].

With the proposed potentials in mind, various solutions to the challenges in HDC are discussed by De Marinis et al.[56], starting with optical interconnects offering higher bandwidth and lower energy consumption to facilitate data transfer between processors and memory. Opting for passive optical components such as optical interference units (OIUs) and optical nonlinear units (ONUs) enables efficient operations with negligible optical power consumption. Hybrid optical-electrical structures take advantage of the speed and efficiency in optical components alongside the control capabilities of electronic elements[56]. In addition, approaches based on time-domain multiplexing have shown to be particularly powerful for the implementation of photonic quantum computing schemes based on continuous variable (CV) encoding[57–59]. Ultra-large cluster states consisting of up to one million of entangled modes multiplexed in the time-domain have been already experimentally demonstrated[60]. Such clusters have been recently exploited for the first realization of a programmable Boson sampler able to show quantum computational advantage, and represent a promising resource for fault-tolerant universal quantum photonic computing[61,62].

## 3. Orthogonal degrees of freedom for photonic computing

In order to satisfy growing computational scaling, hybrid use of orthogonal degrees of freedom in Bosonic computing enables high throughput and parallelization. Using photons, a convenient parameter for analog computing is temporal modulation. Due to the enormous bandwidth of optical carrier frequencies, moving to high modulation frequencies and short pulse durations enables computation with very low latency. At the same time, temporal degrees of freedom can be exploited to minimize the number of required physical components. Demonstrations of low energy consumption photonic computation in edge applications[63] have been realized, utilizing temporal accumulation for large vector representation[8,63,64] (Fig. 3a) and photonic convolutional neural networks[53]. Additionally, large-scale quantum photonic processors[57,65], particularly for cluster-state photonic quantum computation[58,59,66], have been showcased. Temporal multiplexing is typically achieved by introducing time offsets using delay lines[8,65] (Fig. 3b).

A complementary form of multiplexing involves utilizing the wavelength or frequency of the light. As photons do not interact with each other at the power levels required for analog computation, multiple signal carriers can be processed in parallel by encoding individual information on different wavelengths or spectral channels (Fig. 3c). This approach enables parallel processing in linear photonic analog computation[8,48,54] and also allows for creating highly entangled non-classical states for photonic quantum computation[36,67,68].

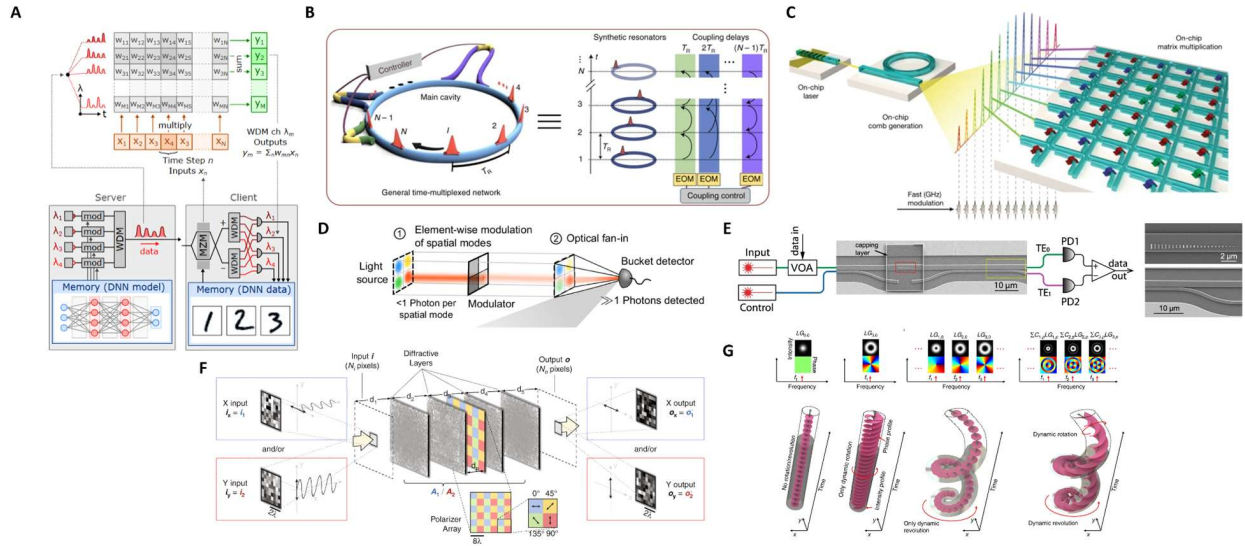

**Figure 3: Degrees of freedom in photonic computing.** A) Schematic of a MVM which is realized by partial summation in the detectors[63] which is making use of temporal multiplexing. B) Example of delay lines for the implementation of temporal multiplexing in a photonic resonator network[65]. C) Implementation of wavelength division multiplexing for an integrated tensor core[48]. D) Scheme of spatial light modulators used for the implementation of weights for optical dot product calculations[71]. E) SEM image of a programmable multi-mode converter using a phase-change material metasurface on the waveguide[74]. F) Layout of an optical polarization-encoded diffractive network[76]. G) Example of the influence of the orbital angular momentum on the beam shape of light[83].

Spatial mode multiplexing with orthogonal free-space or waveguide modes enables parallel processing of analog data in physical systems which support multiple propagation modes. The abundance of spatial modes in free space optics scenarios allows for highly parallel manipulation and processing of large input states. This parallel processing is achieved through wavefront shaping with lenses, metasurfaces[69,70] or spatial light modulators[71,72] (Fig. 3d). In integrated and fiber-based systems, spatial multiplexing is realized by using different waveguides and fibers as separate channels[8,29]. In integrated photonic platforms employed for physical computation, waveguides carrying coherent or incoherent signals replace the traditional electrical wires. Transitioning from single-mode to multi-mode waveguide geometries support the use of higher order modes to encode information in orthogonal states. This approach has been successfully applied in mode-multiplexed matrix-vector multiplication and convolutional neural networks[73,74] (Fig. 3e).

As an intrinsic property of optical waves, polarization has a long-standing record of encoding orthogonal information across various fields, including optical communication applications[75], optical processors[76], quantum optics[77] and quantum information processing[78–80] (Fig. 3f). Lastly, the orbital angular momenta (OAM) carried by light-waves provides additional pathways for information encoding that are not exploited in conventional fermionic computation (Fig. 3g). OAM is predominantly employed in quantum applications[81–85] and free space communication[86], offering unparalleled opportunities to advance data encoding and processing.

## 4. Trajectories and perspectives

Integrating classical photonic processing with quantum photonics opens up possibilities for developing neuromorphic quantum photonics. This fusion leverages the high data throughput of photonic neural networks alongside the capability of quantum networks to process information within a high-dimensional Hilbert space. Since quantum systems can naturally exploit a multidimensional space for physical computing, it makes them exceptionally suitable for implementing tasks which are typical in machine learning and artificial intelligence, such as performing linear operations on large sets of data[88] or mapping data into a high-dimensional feature space[89,90]. Over the last decade, this has motivated the growth of a new research field known as *quantum machine learning*[91]. At present, quantum machine learning can be broadly classified into two main categories, i) Quantum basic linear algebra sub-routines (qBLAS) and ii) Quantum neural networks (QNNs).

qBLAS are quantum algorithms designed to efficiently execute sub-routines based on linear algebra operations, such as inverting linear systems of equations[88]. These algorithms are valuable in various machine learning tasks e.g., least-square data fitting[92] and principal component analysis[93], enhancing computational efficiency and performance in these applications. Using only $n$ logical qubits, qBLAS can process a large dataset with $2^n$– dimensionality, with an exponential speed-up over their best-known classical counterparts. Despite their significant potential, qBLAS are subject to several limitations and challenges[94] that must be addressed. One of the most notable challenges with qBLAS is the efficient loading of $2^n$ data entries into $n$ quantum bits, a task known as the *input problem*, which involves the use of a quantum random access memory[95]. Additionally, the application of qBLAS strictly requires large-scale fault-tolerant quantum computers. Although recent advancements have been made in the realization of fault-tolerant quantum computing architectures[96], such technologies are far from being available in the near future.

QNNs are variational quantum circuits based on a set of parametrized quantum gates[97]. Unlike the case of qBLAS, such algorithms can be implemented on near-intermediate scale quantum (NISQ) processors, which are medium-sized quantum computer, typically consisting of a few hundreds of modes and gates. NISQ processors can demonstrate quantum computational advantage without the need for extensive error correction[98]. When handling classical data sets, QNNs can perform most of the tasks assigned to classical neural networks. They also address the *input problem* described in the previous paragraph by establishing a direct (one-to-one) or near-direct (close to one-to-one) correspondence between input data entries and the physical modes of the network. However, unlike the case of qBLAS, the question of whether these systems can provide a genuine advantage over classical systems remains more nuanced [99,100]. As noted in Ref. [91], there are compelling reasons to believe that such an advantage is possible. Indeed, if a quantum processor can generate outputs that are computationally difficult for classical systems to reproduce, it is also reasonable to assume that it could identify patterns in data that a classical computer might miss. For example, in supervised learning tasks, a first set of quantum gates is employed to map the input data into an exponentially large (but sparse) feature Hilbert space. The mapped data can be directly classified in the feature Hilbert space with an additional set of parametrized quantum gates, whose values - in close analogy with classical neural networks - are optimized to minimize a cost function[101,102]. Preliminary results allude that these systems can enable a higher capacity, i.e., the ability of approximating a larger number of functions, as well as a faster training than their classical counterparts[103]. Interestingly enough, a classical simulation of a QNN consisting of a sizeable number of modes quickly becomes a computationally intractable problem. Thus, a proper benchmarking of such hypotheses, will unavoidably require the physical implementation of these architectures.

Moving forward towards the implementation of QNNs in near-term quantum photonic processors, two main approaches are currently under study: discrete-variable (DV) and continuous-variable (CV) optical quantum computing. In DV quantum computing, the information is encoded in discrete degrees of freedom of a single photon, such as its orthogonal polarization states, or its alternative propagating paths. Challenges ahead include the development of deterministic single-photon sources, as well as deterministic two-photon entangling gates, for which nonlinearities at the single-photon level (currently unavailable with conventional nonlinear optical media) are strictly required. In this direction, a particularly promising approach is the use of quantum dots embedded in waveguides or optical cavities. These systems are already a leading platform for the realization of highly efficient single-photon emitters[104–106] and, as recently demonstrated, can also be employed for mediating nonlinear interactions between single-photon wavepackets[107–110]. Alternatively, within the narrower framework of quantum reservoir computing[111], the required nonlinearity could be achieved through quantum memristors, which can be more easily implemented in current photonic platforms by combining programmable interferometers with single-photon detectors[112].

Moreover, CV quantum optics adopts an analog approach to optical quantum computing, where information is encoded not in single photons but in operators which are CVs, namely, the amplitude and phase quadratures of the electric field[113]. As a main advantage, the required quantum states - squeezed states of light - can be generated on-demand by parametric down-conversion, while deterministic entangling operations can be performed using beam splitters. Additionally, the information encoded in the states is read-out by homodyne detection with conventional photodiodes. Essentially, when implementing this set of operations (known as *Gaussian operations*) CVQNNs can maintain all the benefits of classical photonic systems, including high bandwidth, low-energy consumption, and operation at room temperature. Furthermore, likewise to classical photonic systems, they can be scaled-up to larger dimensions by leveraging either frequency[114] or temporal[115] degrees of freedom. This property, combined with the possibility of generating squeezed light on-demand and to create entanglement deterministically, grants a considerable potential for scalability. Indeed, using only a few optical components interfaced with optical delay lines, it is already possible to generate ultra-large cluster states consisting of up to ∼1 million of entangled modes multiplexed in the time-domain[60,116].

Nevertheless, to realize a CVQNN and demonstrate any quantum computational advantage[117], Gaussian operations alone are insufficient. As with DV systems, strong optical nonlinearities, e.g., a self-Kerr interaction, are required to complement Gaussian operations. These nonlinearities play an analogous role to the nonlinear activation functions employed in classical neural networks, enabling more complex computational capabilities.

Up to this point, reaching sufficiently strong nonlinearities with the available material platforms has been unfeasible. As a workaround, current CV schemes often make use of photon-number-resolving (PNR) detectors, either alongside or instead of homodyne detectors. However, state-of-the-art PNR detectors, typically based on superconducting transition edge sensors, have several significant limitations[118]. Firstly, they must operate at ∼mK temperatures, requiring the use of costly and energy-intensive Helium-4 dilution refrigerators. Secondly, these detectors are relatively slow, with a maximum bandwidth limited to only a few MHz. Lastly, PNR detectors are not an optimal choice for achieving computational universality, restricting their application primarily to a specific subset of problems known as Gaussian Boson sampling[13,14,119].

Remarkably, the recent progress in the performance of photonic integrated circuits implemented in $\chi^{(2)}$-nonlinear materials is now bringing nonlinear optics close to a regime where such strong optical nonlinearities might be within reach[120,121]. Particularly, recent theoretical results suggest that in a CV setting, non-Gaussian operations with a sufficient strength could potentially be implemented using the existing technology[122–126]. Such an achievement could open entirely new perspectives in quantum information science, enabling the realization of a universal CV quantum photonic processor that can operate entirely at room-temperature.

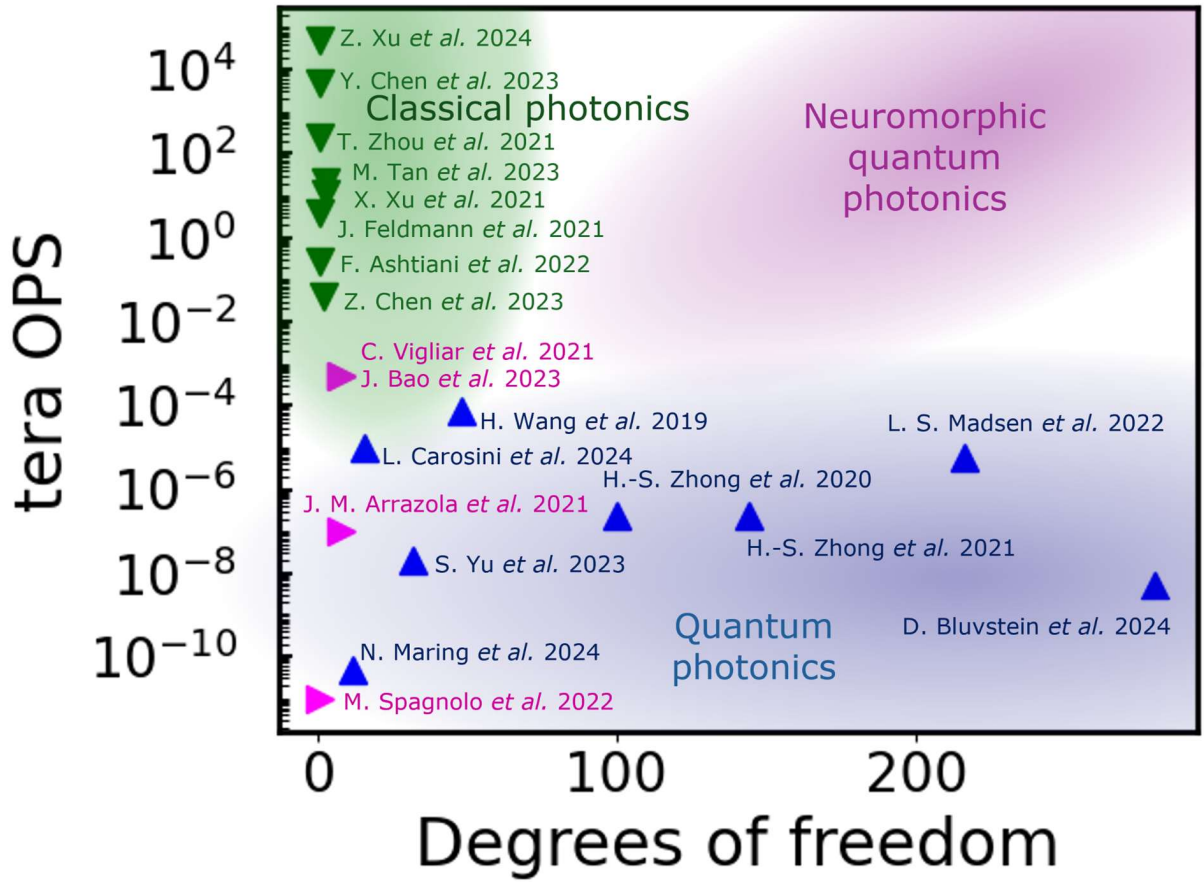

Figure 4: **Perspectives for neuromorphic quantum photonics.** Joining the capabilities of classical photonic processing with quantum photonics opens up a new field of neuromorphic quantum photonics. This combines the high data throughput of photonic neural networks with the ability of quantum networks of processing information in a high-dimensional Hilbert space. References in the graph: [8–12,15,48,50,52,61,62,96,112,127,129–134]

Fig. 4 shows a graphical overview of the current state-of-the art in quantum and classical photonic computing systems. Gaussian boson samplers, which exploit the deterministic generation of squeezed states of light, have demonstrated quantum computational advantage in three distinct experiments[62,127,61]. However, due to the lack of a full circuit programmability, these systems are limited to producing photon number distributions that are computationally difficult to sample classically, without solving any problem of practical relevance. Thus, despite their computational complexity, these systems remain data-wise shallow.

While electronic analog computing has already proven capable of processing large data with low latency and high energy efficiency[128], current photonic processors excel primarily in performing linear operations. Thus, despite their capability of processing large data efficiently, it can be argued that they are computation-wise shallow. CVQNNs offer a potential solution to bridge these two regimes by simultaneously leveraging temporal or frequency degrees of freedom alongside the intrinsic nonlinearity of emerging photonic integrated platforms. This approach enables the development of large-scale optical neural networks that retain the benefits of classical photonic systems while expanding their capabilities, potentially unlocking novel functionalities in artificial intelligence.